\documentstyle[12pt]{article}

\catcode`\@=11
\long\def\@makefntext#1{
\protect\noindent \hbox to 3.2pt {\hskip-.9pt
$^{{\ninerm\@thefnmark}}$\hfil}#1\hfill}                

\def\@makefnmark{\hbox to 0pt{$^{\@thefnmark}$\hss}}  

\def\ps@myheadings{\let\@mkboth\@gobbletwo
\def\@oddhead{\hbox{}
\rightmark\hfil\ninerm\thepage}
\def\@oddfoot{}\def\@evenhead{\ninerm\thepage\hfil
\leftmark\hbox{}}\def\@evenfoot{}
\def\sectionmark##1{}\def\subsectionmark##1{}}

\renewcommand{\thefootnote}{\fnsymbol{footnote}}

\def\sectionc{\@startsection {section}{1}{\z@}{-3.5ex plus -1ex minus 
    -.2ex}{2.3ex plus .2ex}{\bf }}
\def\subsectionc{\@startsection{subsection}{2}{\z@}{-3.25ex plus -1ex minus 
   -.2ex}{1.5ex plus .2ex}{\it }}
\renewcommand{\section}[1]{\sectionc{#1}\hspace*{\parindent}}
\renewcommand{\subsection}[1]{\subsectionc{#1}\hspace*{\parindent}}
\newcounter{appendixc}
\newcounter{subappendixc}[appendixc]
\newcounter{subsubappendixc}[subappendixc]

\renewcommand{\appendix}[1] {\vspace*{0.6cm}
        \refstepcounter{appendixc}
        \setcounter{figure}{0}
        \setcounter{table}{0}
        \setcounter{equation}{0}
        \renewcommand{\thefigure}{\Alph{appendixc}.\arabic{figure}}
        \renewcommand{\thetable}{\Alph{appendixc}.\arabic{table}}
        \renewcommand{\theappendixc}{\Alph{appendixc}}
        \renewcommand{\theequation}{\Alph{appendixc}.\arabic{equation}}
        \noindent{\bf Appendix \theappendixc #1}\par\vspace*{0.4cm}}

\def\abstracts#1{{
        \centering{\begin{minipage}{13.2truecm}
        \footnotesize\baselineskip=13pt\noindent
        \parindent=0pt #1
        \end{minipage}}\par}}

\renewenvironment{thebibliography}[1]
        {\begin{list}{\arabic{enumi}.}
        {\usecounter{enumi}\setlength{\parsep}{0pt}
\setlength{\leftmargin 0.75cm}{\rightmargin 0pt}
         \setlength{\itemsep}{0pt} \settowidth
        {\labelwidth}{#1.}\sloppy}}{\end{list}}

\newcounter{itemlistc}
\newcounter{romanlistc}
\newcounter{alphlistc}
\newcounter{arabiclistc}

\newcommand{\fcaption}[1]{
        \refstepcounter{figure}
        \setbox\@tempboxa = \hbox{\footnotesize Figure~\thefigure. #1}
        \ifdim \wd\@tempboxa > 6in
           {\begin{center}
        \parbox{6in}{\footnotesize\baselineskip=13pt Figure~\thefigure. #1}
            \end{center}}
        \else
             {\begin{center}
             {\footnotesize Figure~\thefigure. #1}
              \end{center}}
        \fi}

\newcommand{\tcaption}[1]{
        \refstepcounter{table}
        \setbox\@tempboxa = \hbox{\footnotesize Table~\thetable. #1}
        \ifdim \wd\@tempboxa > 6in
           {\begin{center}
        \parbox{6in}{\footnotesize\baselineskip=13pt Table~\thetable. #1}
            \end{center}}
        \else
             {\begin{center}
             {\footnotesize Table~\thetable. #1}
              \end{center}}
        \fi}

\def\@citex[#1]#2{\if@filesw\immediate\write\@auxout
        {\string\citation{#2}}\fi
\def\@citea{}\@cite{\@for\@citeb:=#2\do
        {\@citea\def\@citea{,}\@ifundefined
        {b@\@citeb}{{\bf ?}\@warning
        {Citation `\@citeb' on page \thepage \space undefined}}
        {\csname b@\@citeb\endcsname}}}{#1}}

\newif\if@cghi
\def\cite{\@cghitrue\@ifnextchar [{\@tempswatrue
        \@citex}{\@tempswafalse\@citex[]}}
\def\citelow{\@cghifalse\@ifnextchar [{\@tempswatrue
        \@citex}{\@tempswafalse\@citex[]}}
\def\@cite#1#2{{$\null^{#1}$\if@tempswa\typeout
        {IJCGA warning: optional citation argument
        ignored: `#2'} \fi}}

 1
 1
 1

\font\ninerm=cmr9

\def\thefootnote{\fnsymbol{footnote}}
\def\bea {\begin{eqnarray}}
\def\eea {\end{eqnarray}}
\def\be {\begin{equation}}
\def\ee {\end{equation}}
\def\ben{\begin{enumerate}}
\def\een{\end{enumerate}}
\def\bi{\begin{itemize}}
\def\ei{\end{itemize}}
\def\ie{{\it i.e.}}
\def\viz{{\it viz.}\ }

\def\etal{{\it et al.}}

\def\O{{\cal O}}

\def\H{{\cal H}}

\def\prl {{\it Phys. Rev. Lett.\ }}
\def\pl {{\it Phys. Lett.\ }}
\def\pr {{\it Phys. Rev.\ }}
\def\np {{\it Nucl. Phys.\ }}

\def\gL{g_{\mbox{\tiny L}}}
\def\gLeff{g_{\mbox{\tiny L,eff}}}
\def\gS{g_{\mbox{\tiny S}}}
\def\gSeff{g_{\mbox{\tiny S,eff}}}

\def\gP{g_{\mbox{\tiny P}}}
\def\gPeff{g_{\mbox{\tiny P,eff}}}

\def\hyphen{{\mbox{-}}}

\newcommand{\sfrac}[2]{\mbox{\small{$\frac{#1}{#2}$}}}

\begin{document}

\centerline{\normalsize\bf NON-NUCLEONIC EFFECTS STUDIED BY NUCLEAR MOMENTS }
\baselineskip=15pt

\vspace*{0.6cm}
\centerline{\footnotesize I.S. TOWNER}
\baselineskip=13pt
\centerline{\footnotesize\it AECL, Chalk River Laboratories,
Chalk River, Ontario K0J 1J0, Canada}
\baselineskip=13pt
\centerline{\footnotesize E-mail: towneri@cu1.crl.aecl.ca}

\vspace*{0.6cm}
\abstracts{Measured magnetic moments of single-particle states in
the Pb region differ significantly from their Schmidt values.  We
discuss the reasons for this in terms of meson-exchange currents,
isobar currents, core polarisation and other effects.}

\normalsize\baselineskip=15pt
\setcounter{footnote}{0}
\renewcommand{\thefootnote}{\alph{footnote}}

\section{Meson-exchange currents}\label{sec:mec}  
Nuclear physics calculations are based on the nonrelativistic 
Schr\"{o}dinger equation in which wavefunctions for a many-body state
are computed as a first step and expectation values of observables
calculated as a second step.  For processes such as $\gamma$- and
$\beta$-decay the observable is represented by one-body operators,
whose particular form can easily be deduced once the relativistic
interaction between the currents and the fields is written down.
For example, the interaction of a nucleon charge current with an
electromagnetic field is given by a Hamiltonian $\H = - J_{\mu} A_{\mu}$
where $A_{\mu}$ is the vector potential describing the field and
$J_{\mu}$ is the charge current of a single nucleon:

\be
J_{\mu}(k) = {\rm i} \overline{u}(p^{\prime})
\left [ F_1 \gamma_{\mu} - \frac{F_2}{2M} \sigma_{\mu \nu} k_{\nu}
\right ] u(p) ,
\label{eq:jmu}
\ee

\noindent where $k = p^{\prime} - p$.
Here $\overline{u}(p^{\prime})$ and $u(p)$ are plane-wave Dirac
spinors, and
$F_1$ and $F_2$ are the Dirac and
anomalous coupling constants.  A nonrelativistic form of the charge
current $J_{\mu} = (\rho, {\bf J})$ is obtained by
multiplying out the Dirac spinors with the Dirac matrices and
keeping only terms to leading order in $1/M$ --- this produces
a one-body operator sandwiched between Pauli spinors:

\bea
\rho(k) & = & F_1 + \O(1/M^2) 
\nonumber \\
{\bf J}(k) & = & \frac{F_1}{2M} ({\bf p}^{\prime}
+ {\bf p}) + \frac{F_1 + F_2}{2M} {\rm i}\mbox{\boldmath $\sigma$}
\times {\bf k} + \O(1/M^3) ,
\label{eq:rJ}
\eea

\noindent where for economy the Pauli spinors associated with
$\overline{u}
(p^{\prime})$ and $u(p)$ have been omitted.   A Fourier transform
to coordinate space and a multipole decomposition\cite{CT90} 
leads to
the familiar one-body electromagnetic operators, whose matrix
elements are then evaluated in the many-body system using, for example,
shell-model wavefunctions.  When the coupling constants for the
nucleon current embedded in the nuclear medium
are taken from
the free-nucleon system then the procedure is called the {\it impulse
approximation}.  It has been widely successful, but at some point
it is known to break down.  This is because the nucleons in the nucleus
are interacting through the exchange of mesons and the perturbing
electromagnetic field can disturb this exchange and even interact
with the exchanged meson itself.  As has been dramatically shown
by Kubodera, Delorme and Rho\cite{KDR78} with soft-pion theorems,
the most important meson-exchange corrections (MEC) occur in
magnetic moments and transitions (operators originating in the
space component of the vector current, $J_{\mu}$) and in axial-charge
transitions in beta decay (operators originating in the time
part of the axial current, $J_{5 \mu}$).

There are different approaches to the construction of the two-body
MEC operators that are equivalent up to certain unitary
transformations\cite{AGH93}.  They are: {\it (i)} the 
quasipotential formalism
in which one starts from the covariant four-dimensional Bethe-Salpeter
equation and uses the Blankenbecler-Sugar reduction to get a
three-dimensional equation\cite{JW87}; {\it (ii)} the equations
of motion method in which unwanted degrees of freedom are
eliminated by a unitary FST transformation\cite{FST54}, and the
nonrelativistic reduction effected by a Foldy-Wouthuysen
transformation\cite{Fr77,HG76,GA92,Ta92}; and {\it (iii)} the extended
S-matrix formalism in which the S-matrix is expanded in a series of
Feynman diagrams\cite{CR71,ATA89}.  Our calculations are based on this 
latter approach using a phenomenological chiral Lagrangian for
the isovector mesons, $\pi$, $\rho$, and $A_1$,\cite{AT84} and
augmented by the standard isoscalar meson Lagrangians for
$\sigma$ and $\omega$ mesons.

\subsection{`External' corrections} \label{subsec:exc}
There are two types of Feynman diagrams to consider: {\it (i)} those
in which the electromagnetic field interacts with the nucleon
charge current, $J_{\mu}(k)$,
to be known generically as ``pair" graphs; and {\it (ii)}
those in which the electromagnetic field interacts with the
exchanged meson's charge current, to
be known as ``current" graphs.  There is one comment we wish to
make concerning the construction of the ``pair" graph.  The
general form of its expression involves the product

\be
J_{\mu}(k) S(Q) \Gamma(-q) ,
\label{eq:JSG}
\ee

\noindent where $J_{\mu}(k)$ is the nucleon charge current, Eq.~(\ref{eq:jmu}),
$S(Q)$ is the nucleon propagator of momentum $Q$,
and $\Gamma (-q)$ is the meson-production vertex of momentum $q$. 
To avoid double counting, the contribution
from the standard one-nucleon current must be carefully identified
and separated from the ``pair" graph leaving the genuine two-nucleon
MEC operators.  This separation is usually done by dividing the
nucleon propagator into positive and negative frequency components

\be
S(Q) = S^{(+)}(Q) + S^{(-)}(Q)
\label{eq:SQ}
\ee

\noindent and retaining only the negative frequency component, 
$S^{(-)}(Q)$.  This gives the correct result providing $J_{\mu}(k)$
and $\Gamma(-q)$ are energy independent, \ie \ $J_{\mu}(k)$ does
not depend on $k_0$ and $\Gamma(-q)$ does not depend on $q_0$.
However, for the magnetic moment, which derives from the spatial
component of $J_{\mu}(k)$, Eq.~(\ref{eq:jmu}), 
there is a dependence on $k_0$ coming
from the anomalous term $(F_2/2M) \sigma_{i 0} k_0$.  In this
case Adam \etal \cite{ATA89} define

\be
J_{\mu}(k) = \overline{J}_{\mu}(k) + k_0 J_{\mu}^{\prime}(k) ,
\label{eq:jov}
\ee

\noindent where $\overline{J}_{\mu}(k)$ has no dependence on $k_0$.
Thus $J_{\mu}^{\prime}(k) = \partial /(\partial k_0) J_{\mu}(k)$.
Similarly defining

\be
S^{(+)}(Q) = \frac{1}{k_0} S^{\prime (+)}(Q)
\label{eq:spr}
\ee

\noindent then an additional term from the ``pair" graph that
contributes a genuine MEC involves the combination

\be
J_{\mu}^{\prime}(k) S^{\prime (+)}(Q) \Gamma (-q) .
\label{eq:ext}
\ee

\noindent Such corrections are called `external' by Adam \etal
\cite{ATA89}.  Similarly when $\Gamma (-q)$ depends on $q_0$ there
are `vertex' corrections.  The external corrections contribute
to leading order, $\O (1/M^2)$, for scalar and vector meson
exchanges, and to next-to-leading order in $\O (1/M^4)$ for
pseudoscalar meson exchanges.  Here $M$ is the nucleon mass.
Vertex corrections only contribute to next-to-leading order.
We take this opportunity to add the `external' corrections
to our previously computed\cite{To87} MEC corrections for magnetic 
moments in the Pb region.  The correction does not have a large
impact, being of order $\sim \pm 0.08 \mu_{N}$ with a plus
sign for a single-particle proton state with $j = \ell + \sfrac{1}{2}$
or a neutron with $j = \ell - \sfrac{1}{2}$, and a minus sign in
the other two cases.

\subsection{MEC corrections for states in Pb region} \label{subsec:pb}
The procedure for constructing the two-body MEC operators for
a magnetic moment calculation is as follows\cite{To87}:

\bi
\item Write down expressions for the ``pair" and ``current" Feynman
diagrams in terms of Dirac spinors for the nucleons for the case
when the Lorentz index on the charge current, $J_{\mu}(k)$, is
space-like;
\item Expand the Dirac spinors in powers of $(1/M)$ and retain
leading terms, to be denoted ${\bf J}(k)$;
\item Construct the magnetic moment operator $\mbox{\boldmath $\mu$}
= - \sfrac{1}{2} {\rm i} \mbox{\boldmath $\nabla$}_{k} \times {\bf J}(k) 
\mid_{k \rightarrow 0}$;
\item Fourier transform to coordinate space to obtain the two-body
operator, $\mbox{\boldmath $\mu$}({\bf r,R})$,
where ${\bf r}  =  {\bf r}_1 -  {\bf r}_2$ 
and ${\bf R} = \sfrac{1}{2}
({\bf r}_1 +  {\bf r}_2 )$ 
are relative and centre of mass coordinates;
\item Compute matrix elements of this operator in a many-body system
using, for example, nuclear wavefunctions from a shell-model
calculation.
\ei

We can simplify the last stage by only considering matrix elements of
the magnetic moment operator in closed-shell-plus (or minus)-one
configurations.  For the impulse-approximation one-body operator the
calculation reduces simply to the single-particle matrix element
for the valence orbital outside the closed shell (the Schmidt value).
Recall that the impulse-approximation magnetic-moment operator is

\be
\mbox{\boldmath $\mu$} = \gL {\bf L} + \gS {\bf S}
\label{eq:CP3}
\ee

\noindent with $\gL = 1$ and $\gS = 5.586$ for a proton, and
$\gL = 0$ and $\gS = -3.826$ for a neutron.  The magnetic-moment
expectation value for a single particle of angular momentum $a$ is then

\be
\mu = \left ( \frac{a}{a+1} \right )^{1/2}
\langle a \| \gL {\bf L} + \gS {\bf S} \| a \rangle
\label{eq:CP4}
\ee

\noindent The reduced matrix elements are defined in the conventions of
Brink and Satchler\cite{BS68}.  
For the two-body MEC operators, the calculation becomes one of
computing two-body matrix elements between the valence nucleon and one
of the nucleons in the closed-shell core and summing over all
the nucleons in the core.  It is useful to express the result of this
computation in terms of an equivalent one-body operator

\be
\mbox{\boldmath $\mu$}_{\mbox{\tiny eff}} = \gLeff {\bf L}
+ \gSeff {\bf S} + \gPeff [Y_2,{\bf S}] ,
\label{eq:gef}
\ee

\noindent where $\gLeff = \gL + \delta \gL $ etc..  Here $\gL$ is
the free-nucleon coupling constant of Eq.~(\ref{eq:CP3}) 
and $\delta \gL$ the calculated
correction.

\begin{table}[t]
\protect
\tcaption{ Meson-exchange current corrections to magnetic moments
in $^{209}$Bi and $^{209}$Pb.}
\label{tab:tabl18}  
\small
\vspace{0.4cm}
\begin{center}
\begin{tabular}{lrrrrcrrrr}
\hline \\[-3mm]
 & \multicolumn{4}{c}{Proton $0h_{9/2}$} &~& 
 \multicolumn{4}{c}{Neutron $0i_{13/2}$} \\ 
\cline{2-5}
\cline{7-10}
 & & & & & & & & & \\[-3mm]
 & $ \delta \gL$ & $ \delta \gS$ & $\delta \gP$ & $\Delta \mu$ &
 & $ \delta \gL$ & $ \delta \gS$ & $\delta \gP$ & $\Delta \mu$ \\
 & & & & & & & & & \\[-3mm]
\hline \\[-3mm]
$\pi$-pair & 0.303 & 0.725 & 0.302 & 1.16 & & $-0.168$ & $-0.336 $ &
0.004 & $-1.18$ \\
$\pi$-current & $-0.247$ & $-0.384$ & $-1.048$ & $-0.94$ & & 0.136 &
0.215 & 1.164 & 1.02 \\
$\rho \hyphen \pi$ & $-0.000$ & $-0.002$ & 0.019 & $-0.00$ & & $-0.000$ &
$-0.001$ & 0.017 & 0.00 \\
$\omega \hyphen \pi$ & $-0.000$ & $-0.031$ & 0.151 & $-0.01$ & & 0.000 &
0.026 & $-0.190$ & 0.00 \\
A$_{1} \hyphen \pi$ & 0.000 & 0.002 & 0.019 & $-0.00$ & & $-0.000$ &
$-0.001$ & $-0.014$ & $-0.00$ \\
$\sigma$-pair & 0.033 & 0.227 & 0.500 & 0.02 & & 0.002 & $-0.186$ &
$-0.560$ & $-0.13$  \\
$\omega$-pair & $-0.004$ & $-0.080$ & $-0.244$ & 0.04 & & $-0.004$ &
0.071 & 0.261 & 0.03 \\
$\rho$-pair & 0.078 & 0.072 & $-0.057$ & 0.36 & & $-0.042$ &
$-0.009$ & 0.046 & $-0.25$ \\
$\rho$-current & 0.036 & 0.046 & 0.029 & 0.16 & & $-0.020$ &
$-0.013$ & $-0.019$ & $-0.13$ \\     
A$_{1}$-current & $-0.013$ & $-0.006$ & 0.007 & $-0.06$ & & 0.007 &
0.003 & $-0.007$ & 0.04 \\[3mm]
~~~~Total & 0.186 & 0.560 & $-0.346$ & 0.72 & & $-0.089$ &
$-0.226$ & 0.713 & $-0.59$ \\[3mm]
\hline
\end{tabular}
\end{center}
\end{table}

In Table \ref{tab:tabl18} we give some sample results for magnetic
moments for a valence $0h_{9/2}$ proton and $0i_{13/2}$ neutron outside
a $^{208}$Pb closed-shell core.  Harmonic oscillator radial 
wavefunctions, $\hbar \omega = 7.0$ MeV, are used.  To accommodate
the role of short-range correlations in the nuclear medium, we have
multiplied the two-body MEC operator by a correlation function
$g(r)$ chosen to be $g(r) = \theta (r - d)$ with $d = 0.7$ fm. We
also include vertex form factors at all meson-nucleon vertices,
taking care to ensure that the equation of continuity remains
satisfied\cite{Ri84,Ma84}.

Let us look at the $\delta \gL$ value.  In the S-matrix
approach, the pion-pair and pion-current diagrams give large
contributions but there is a significant cancellation between
them.  For example, for the $h_{9/2}$ proton at $^{209}$Bi we compute
$\delta \gL = 0.303 - 0.247 = 0.056$ with vertex form factors and
$\delta \gL = 0.325 - 0.230 = 0.095$ without.  This latter value is
very close to the estimate first obtained by Miyazawa\cite{Mi51}
of $\delta \gL \simeq 0.1$ from pion-exchange currents.  We find,
however, that heavy mesons make a substantial contribution to
$\delta \gL$. Our result in Table \ref{tab:tabl18} is
$\delta \gL = 0.056 + 0.130 = 0.186$ for a $0h_{9/2}$ proton and   
$\delta \gL = -0.032 - 0.057 = -0.089$ for a $0i_{13/2}$ neutron,  
where the first figure represents contributions from pions, the 
second from heavy mesons. As a consequence, the magnetic-moment
correction from MEC is positive for protons and negative for
neutrons.  

The first experimental indication that the proton $\gL$ value
is enhanced in nuclei by about 10\% over its free-nucleon value
came in the measurement of Yamazaki \etal \cite{Ya70} of the
magnetic moment of the $11^{-}$ isomer in $^{210}$Po.
Nagamiya and Yamazaki\cite{NY71} later showed the enhancement
to be a general phenomenon present throughout the whole mass
region.  A systematic analysis\cite{Ya79} of all magnetic
moment data in the Pb region produces as best-fit values:
$\delta \gL = 0.15 \pm 0.02$ for a proton, and
$\delta \gL = -0.03 \pm 0.02$ for a neutron.  Our results in
Table \ref{tab:tabl18} are in reasonable accord with this expectation,
but there are other corrections, core polarisation, isobar currents
etc., to be considered before a serious comparison with
experimental data can be made.

\subsection{Isobar currents} \label{subsec:ic}
The MEC currents constructed so far involve nucleons interacting with
mesons.  There is a further process to be considered, in which the meson
prompts the nucleon to be raised to an excited state, the 
$\Delta$-isobar resonance, which is then deexcited by the
electromagnetic field.  Such proceses as these are called `Isobar
currents'.  
In Table \ref{tab:tabl29} it is seen that,
the isobar current mainly quenches the spin $\gS$-factor by about 10\%
over its free-nucleon value producing a magnetic-moment correction
that is positive for protons in $j = \ell + \sfrac{1}{2}$ orbits
and neutrons in $j = \ell - \sfrac{1}{2}$ orbits, and negative otherwise.

\section{Core polarisation} \label{sec:corepol}
The nuclear wave function in a nucleus such as $^{209}$Bi 
is more complicated than a single particle
coupled to a closed-shell core and will have components of
$2p$-$1h$ and $3p$-$2h$ configuration whose impact on the magnetic-moment
expectation value can be estimated in perturbation theory.  To first
order in the residual interaction, $V$, the only contributions arise
from particle-hole states coupled to the same angular momentum as the
multipolarity of the operator, $\lambda$; namely, $\lambda = 1$ for
magnetic moments.  At the $^{208}$Pb closed shell, there are two 
such particle-hole states: proton $(h^{-1}_{11/2} h_{9/2})$ and
neutron $(i^{-1}_{13/2} i_{11/2})$, and their contribution to
the magnetic moment of $^{209}$Bi to first- and second-order
is given by the expression\cite{To87}:

\be
\langle a \| \Delta \mbox{\boldmath $\mu$} \| a \rangle
= 2 \sum_{\alpha} T_{\alpha} \frac{L_{\alpha}}{\epsilon_{\alpha}}
+ 2 \sum_{\alpha \beta} T_{\alpha} \frac{(A-B)_{\alpha \beta}}
{\epsilon_{\alpha}} \frac{L_{\beta}}{\epsilon_{\beta}}
\label{eq:CP1}
\ee

\noindent where Greek letters represent the particle-hole coupled 
states, \viz $\mid \alpha \rangle = $  
$ \mid (h^{-1}_{\alpha} p_{\alpha} )
\lambda \rangle $ of multipolarity $\lambda$ and the following notation
is introduced:

\bea
T_{\alpha} & = & \langle 0 \| \mbox{\boldmath $\mu$} \|
(h^{-1}_{\alpha} p_{\alpha} ) \lambda \rangle
\nonumber \\
L_{\alpha} & = & - \hat{a}^{-1} \langle (h^{-1}_{\alpha} p_{\alpha}) 
\lambda \mid V \mid (a^{-1} a) \lambda \rangle 
\nonumber \\
A_{\alpha \beta} & = & \langle (h^{-1}_{\alpha} p_{\alpha}) \lambda \mid
V \mid (h^{-1}_{\beta} p_{\beta} ) \lambda \rangle
\nonumber \\
B_{\alpha \beta} & = & \langle 0 \mid V \mid (h^{-1}_{\alpha} p_{\alpha} )
\lambda , (h^{-1}_{\beta} p_{\beta}) \lambda \rangle
\nonumber \\
& = & (-)^{h_{\alpha} - p_{\alpha} + \lambda } \langle (p^{-1}_{\alpha}
h_{\alpha }) \lambda \mid V \mid (h^{-1}_{\beta} p_{\beta}) \lambda 
\rangle
\nonumber \\
\epsilon_{\alpha} & = & \epsilon_{h_{\alpha}} - \epsilon_{p_{\alpha}}
\label{eq:CP2}
\eea

\noindent with $\hat{a} = (2 a + 1)^{1/2}$.  Here $a$ 
represents the valence nucleon outside the closed-shell core.  The
energy denominators, $\epsilon_{\alpha}$, are estimated from known 
spin-orbit splittings of $-5.6$ MeV for the proton $h$-orbitals
and $-5.85$ MeV for the neutron $i$-orbitals.
The residual interaction, $V$, is taken
as a one-boson-exchange potential involving $\pi$, $\rho$, $\omega$,
$\sigma$ and $A_1$ mesons and, for use in finite nuclei, multiplied
by a short-range correlation function as described by 
Towner\cite{To87}.   

The first- and second-terms in $V$ given in Eq.~(\ref{eq:CP1}) can easily
be extended to higher orders in the random phase approximation (RPA).
It must be stressed that these are not the only second-order terms;
there are others that we will discuss further in Sec.~\ref{sec:HO}.
However, these terms are the simplest to evaluate.  The selection rule
that limited the number of particle-hole states, $\alpha$, to just
two still applies to all orders in RPA.  The generalization of
Eq.~(\ref{eq:CP1}) is now

\be
\langle a \| \Delta \mbox{\boldmath $\mu$} \| a \rangle =
2 \sum_{\alpha \beta} T_{\alpha} \left [ I - (A-B)/\epsilon
\right ]^{-1}_{\alpha \beta} L_{\beta} / \epsilon_{\beta}
\label{eq:CP5}
\ee

\noindent where the matrix $\left [ I - (A-B)/\epsilon \right ]$
is first constructed and then inverted.  Here $I$ is the unit matrix.

\begin{table}[t]
\protect
\tcaption{ Contributions to the equivalent effective one-body    
operator from all sources in the OBEP model for a $0h$-proton
and a $0i$-neutron in the Pb region.} \label{tab:tabl29} 
\small
\vspace{0.4cm}
\begin{center}
\begin{tabular}{lrrrrcrrrr}
\hline \\[-3mm]
 & \multicolumn{4}{c}{Proton $0h_{9/2}$} &~
 & \multicolumn{4}{c}{Neutron $0i_{13/2}$} \\ 
\cline{2-5}
\cline{7-10}
 & & & & & & & & & \\[-3mm]
 & $ \delta \gL$ & $ \delta \gS$ & $\delta \gP$ & $\Delta \mu$ &
 & $ \delta \gL$ & $ \delta \gS$ & $\delta \gP$ & $\Delta \mu$ \\
 & & & & & & & & & \\[-3mm]
\hline \\[-3mm]
MEC           & 0.186 & 0.560 & $-0.346$ & 0.72 & & $-0.089$ &
$-0.226$ & 0.713 & $-0.59$ \\
Isobars & $-0.003$ & $-0.452$ & 0.484 & 0.12 & & 0.002 &
0.402 & $-0.680$ & 0.16 \\
CP(RPA)    & 0.005 & $-1.167$ & 0.481 & 0.45 & & $-0.005$ & 1.034 &
0.102 & 0.50 \\
Vib & $-0.030$ & $-0.338$ &  & $-0.01$ & & 0.010 & 0.118 &
         & 0.12  \\
Rel & $-0.024$ & $-0.152$ & $-0.041$ & $-0.05$ & & 0.000 &
0.107 & 0.000 & 0.05 \\
CP(2nd)$^{a}$ & $-0.150$ & $-1.030$ &       & $-0.32$ & & 0.080 &
0.350 &          & 0.66 \\
MEC-CP$^{b}$ & 0.122 & 0.352 &       & 0.46 & & $-0.062$ &
$-0.176$ &          & $-0.46$ \\[3mm]
~~~~Total & 0.106 & $-2.227$ & 0.578 & 1.36 & & $-0.064$ &
1.609 & 0.135 & 0.44 \\[3mm]
\hline \\
\multicolumn{10}{l}{{\footnotesize 
$^{a}$From ref.\cite{AH79}~~~~~~$^{b}$From
ref.\cite{HAS80} } }
\end{tabular}
\end{center}
\end{table}

In Table \ref{tab:tabl29} we give some results
for the core-polarisation calculation, based on Eq.~(\ref{eq:CP5})
and denoted CP(RPA). 
The calculation
based on Eq.~(\ref{eq:CP1}) leads to an alternating sign series:
quenching in first order, enhancement in second order and so on.
As a consequence, the RPA sum is between 30 and 40\% less than
the first-order calculation alone.  The calculation mainly impacts
on the spin $\gS$-factor reducing it by about 25\% compared
to the free-nucleon value.  The magnetic-moment correction is
positive for protons in $j = \ell - \sfrac{1}{2}$ and neutrons in
$j = \ell + \sfrac{1}{2}$ orbits, and negative otherwise.

\subsection{Coupling to core-vibrational states} \label{subsec:vib}
For single-particle states in the Pb region, especially the
high-spin states, their wavefunctions are not those of pure 
single-particle configurations, but have some admixtures with
low-lying vibrational states in the neighbouring even-even nucleus.
For example, the lowest $13/2^{+}$ state in $^{209}$Bi is not
simply a proton $i_{13/2}$ state but has a sizeable admixture
of $h_{9/2} \times 3^{-}$.  Hamamoto\cite{Ha76} has estimated the impact
of core excitations on the magnetic-moment expectation values
in a particle-vibration model.  The correction, denoted `Vib'
in Table~\ref{tab:tabl29}, 
is small, $\leq 0.06 \mu_{N}$,
in all cases except the proton $i_{13/2}$ orbital and,
to a lesser extent, the neutron $i_{13/2}$ orbital.

\subsection{Relativistic correction} \label{subsec:rc}
In deriving the MEC operators, a nonrelativistic reduction is made
of the meson-nucleon interaction.  This reduction is in essence an
expansion in terms of $p/M = v/c$, where $p$ is the typical
nucleon momentum and $M$ its mass.  Terms to
order $(p/M)^2$ were
kept.  The standard one-body operator, Eq.~(\ref{eq:CP3}), has only
been determined to order $(p/M)$ and we will consider its extension
to the next-to-leading order,
$\O (p/M)^3$.  The magnetic moment operator then
becomes\cite{To87}

\be
\mbox{\boldmath $\mu$} = \gL \left \{ {\bf L}
\left ( 1 - \frac{p^2}{2M^2} \right )
- \frac{p^2}{2M^2} \left ( {\bf S} - \left ( {\bf S} \cdot
\hat{{\bf p}} \right ) \hat{{\bf p}} \right ) \right \}
+ \gS {\bf S} \left ( 1 - \frac{p^2}{2M^2} \right )
\label{eq:MEC1}
\ee

\noindent It remains to estimate the expectation value of
$\langle p^2/2M^2 \rangle $ 
for which we use harmonic oscillator wavefunctions.
Since a second derivative is involved, this expectation
value is very sensitive to the choice of radial
wavefunction.  The main result, denoted by `Rel' in Table \ref{tab:tabl29},
is a small reduction in the calculated magnetic moment of
less than 5\%.

\section{Higher orders} \label{sec:HO} 
There are other second-order core-polarization corrections not
contained in the RPA series discussed in Sec.~\ref{sec:corepol}.
Typically these terms have an operator structure, 
$V \mbox{\boldmath $\mu$} V$, where the residual interaction first
excites a single-particle state to a $2p$-$1h$ or $3p$-$2h$
intermediate state where the one-body magnetic-moment 
operator acts, then a
second interaction reduces the intermediate state back to a
single-particle state.  There are no
selection rules to limit the number of intermediate states,
so the calculations are computationally time consuming and have
only been completed without approximation in light nuclei.
In heavy nuclei, Shimizu\cite{Sh76} has explored a closure 
approximation to estimate these terms.

There is another set of terms of the same order, namely first order 
in $V$ but fourth order in meson-nucleon couplings, that have
an operator structure $V {\bf M} + {\bf M} V$, where ${\bf M}$ is
the two-body meson-exchange current operator.  Again there are
no restrictions on the intermediate-state summations.  The
importance of these terms was first pointed out by Arima \etal   
\cite{AH79,HAS80}, where they are known as the `crossing terms'.
Their contribution to the magnetic moment is of 
opposite sign to that from second-order core-polarization terms
involving one-body operators and cancels a large part of it.

\begin{table}[t]
\protect
\tcaption{ Sum of all corrections to magnetic moments and $B(M1)$
in the Pb region (last line gives $\pi g_{7/2}$ case at $^{132}$Sn)
in the OBEP model compared with experiment.}\label{tab:tabl30} 
\small
\vspace{0.4cm}
\begin{center}
\begin{tabular}{lrrrrrrl}
\hline \\[-3mm]
 & \multicolumn{7}{c}{$\Delta \mu$ or $\Delta [ B(M1;i \rightarrow f)
]^{1/2}$ } \\[1mm]
\cline{2-8}
 & & & & & & & \\[-3mm]
 & CP(RPA) & MEC & Isobars & Other & Sum & ASBH$^{a}$ & ~~~~Expt$^{b}$ \\
 & & & & & & & \\[-3mm]
\hline \\[-3mm]
$\pi h_{9/2}$ & 0.45 & 0.72 & 0.12 & 0.08 & 1.36 & 1.49 & ~~\,1.49(0) \\
$\pi i_{13/2}$ & $-0.44$ & 1.07 & $-0.17$ & $-1.21$ & $-0.75$ & $-0.89$ &
$-0.78(10)^{c}$ \\
$\pi s_{1/2}^{-1}$ & $-0.45$ & 0.32 & $-0.29$ & $-0.44$ & $-0.86$ &
$-0.99$ & $-0.92(0)$ \\
$\pi d_{3/2}^{-1}$ & 0.21 & 0.18 & 0.10 & 0.18 & 0.67 & 0.65 
& ~~\,0.64(19)$^{d}$  \\
$\pi f_{5/2} \rightarrow f_{7/2}$ & $-0.37$ & 0.12 & $-0.21$ & $-0.18$
& $-0.63$  & & $-0.82(9)^{e}$ \\[3mm]
$\nu g_{9/2}$ & 0.38 & $-0.37$ & 0.19 & 0.17 & 0.38 & 0.47 
& ~~\,0.44(0) \\
$\nu p_{1/2}^{-1}$ & $-0.09$ & $-0.04$ & $-0.01$ & $-0.05$ &
$-0.18$ & $-0.18$ & $-0.05(0)$ \\
$\nu f_{5/2}^{-1}$ & $-0.30$ & $-0.23$ & $-0.11$ & $-0.03$ & $-0.66$ &
$-0.71$ & $-0.58(3)$ \\
$\nu i_{13/2}^{-1}$ & 0.50 & $-0.59$ & 0.16 & 0.37 & 0.44 
& 0.75 & ~~\,0.90(3)$^{f}$ \\
$\nu p_{3/2}^{-1} \rightarrow p_{1/2}^{-1} $ & $-0.23$ & 0.04 &
$-0.17$ & $-0.06$ & $-0.42$ & & $-0.44(5)^{e}$ \\
$\nu f_{7/2}^{-1} \rightarrow f_{5/2}^{-1}$ & $-0.30$ & 0.06 &
$-0.18$ & $-0.02$ & $-0.44$ & & $-0.52(11)^{e}$ \\[3mm]
$\pi g_{7/2}$ & 0.48 & 0.52 & 0.08 & 0.13 & 1.21 & & ~~\,1.28(1)$^{g}$ \\
\hline \\
\multicolumn{8}{l}{{\footnotesize $^{a}$From ref.\cite{ASBH87} 
~~~~~~ $^{b}$From ref.\cite{Ra89} } } \\
\multicolumn{8}{l}{{\footnotesize $^{c}$Deduced from the magnetic moment of 
the $11^{-}$ state in $^{210}$Po, ref.\cite{Ya70} } } \\
\multicolumn{8}{l}{{\footnotesize $^{d}$Deduced from the magnetic moment of 
the $\sfrac{3}{2}^{+}$ state in $^{205}$Tl, ref.\cite{Ha83} } } \\
\multicolumn{8}{l}{{\footnotesize $^{e}$From ref.\cite{Ha72} } } \\
\multicolumn{8}{l}{{\footnotesize $^{f}$Deduced from the magnetic moment of 
the $12^{+}$ state in $^{206}$Pb, ref.\cite{Na72} } } \\
\multicolumn{8}{l}{{\footnotesize $^{g}$From ref.\cite{St96} } } 
\end{tabular}
\end{center}
\end{table}

In Table \ref{tab:tabl29}, we quote the results of Arima \etal  
\cite{AH79,HAS80}, where the 
cancellation between the second-order core polarisation, CP(2nd),
and the `crossing' term, MEC-CP, is clearly evident,
but in detail it varies from case to case.
The quoted numbers are model dependent and parameter dependent.

 Finally, we bring together all the computed corrections from
core polarisation, CP(RPA), meson-exchange currents, isobar
excitations and together under `Other', the coupling to 
core-vibrational states, the relativistic correction and the
higher-order processes, CP(2nd) and MEC-CP.  In Table \ref{tab:tabl29} 
the results are expressed in terms of an equivalent effective one-body
operator.  Consider first the orbital $\delta \gL$ value.  Our
results of $\delta \gL = 0.106$ for protons and $\delta \gL = -0.064$
for neutrons are not too far from the empirical values determined
in a best-fit analysis of magnetic-moment data in the Pb region
of Yamazaki \etal \cite{Ya79} of $0.15 \pm 0.02$ and
$-0.03 \pm 0.02$ respectively.  The spin $\delta \gS$ values
are dominated by core polarisation.  Our results of
$\delta \gS / \gS = -40 \%$ for protons and $-42 \%$ for neutrons
show a large quenching consistent with an empirical relation
often used: $\gSeff = 0.6 \gS$.  In Table \ref{tab:tabl30} we
compare these corrections with all known magnetic moment and
M1-transition data in closed-shell-plus (or minus)-one nuclei
in the Pb region.  We also list the calculated results obtained
by Arima, Shimizu, Bentz and Hyuga (ASBH)\cite{ASBH87} from
core-polarisation and meson-exchange current processes.  
The agreement between theory and experiment is very good and is
within $0.15 \mu_{N}$ in all cases except one, the neutron $i_{13/2}$
state in $^{207}$Pb, where the strong cancellation between core
polarisation and MEC seems to be too severe.  The ASBH calculation
has less of a cancellation here.

Lastly, in Table \ref{tab:tabl30} we have included one result for the
$^{132}$Sn doubly-closed shell core.  The magnetic moment of
$^{133}$Sb has recently been measured at the OSIRIS mass separator
of the Uppsala University Neutron Physics Laboratory
by Stone \etal \cite{St96}.  In terms of the calculated effective
$g$-factors the results are very similar to $^{209}$Bi.  The computed
magnetic moment is in good agreement with experiment.

\end{document}